\documentclass[11pt,letterpaper]{article}
\usepackage{t1enc}
\usepackage[latin2]{inputenc}
\usepackage{amsmath}
\usepackage{amsfonts}
\usepackage{amssymb}
\usepackage{amsthm}
\usepackage{graphicx}
\usepackage{color}
\usepackage{array}
\usepackage[english,magyar]{babel}
\title{Maximal and minimal realizations of reaction kinetic systems: computation and properties}
\author{Gábor Szederkényi$^1$, Katalin M. Hangos$^1$, Tamás Péni$^2$}
\newenvironment{gmatrix}[1]{\left[\begin{array}[c]{#1}}{\end{array}\right]}

\newtheorem{theorem}{Theorem}[section]

\theoremstyle{definition}

\begin{document}
\selectlanguage{english} \maketitle
\begin{center}
$^{1}$
Process Control Research Group\\
$^{2}$
Systems and Control Laboratory

\medskip
\noindent Computer and Automation Research Institute,\\
Hungarian Academy of Sciences\\
H-1518, P.O. Box 63, Budapest, Hungary\\
Tel: +36 1 279 6000\\
Fax: +36 1 466 7503\\
e-mail: szeder@sztaki.hu\\
\footnotetext[1]{Process Control Research Group, Computer and Automation Research 
Institute, Hungarian Academy of Sciences, Budapest, Hungary, H-1518, P.O. Box 63}
\end{center}

\begin{abstract}
This paper presents new results about the optimization based generation of chemical reaction networks (CRNs) of higher deficiency. Firstly, it is shown that the graph structure of the realization containing the maximal number of reactions is unique if the set of possible complexes is fixed. Secondly, a mixed integer programming based numerical procedure is given for computing a realization containing the minimal/maximal number of complexes. Moreover, the linear inequalities corresponding to full reversibility of the CRN realization are also described. The theoretical results are illustrated on meaningful examples.
\end{abstract}
\newpage

\section{Introduction: chemical reaction networks and their significance}
Positive (nonnegative) systems are characterized by the property that all state variables remain positive (nonnegative) if the trajectories start in the positive (nonnegative) orthant. Thus, positive systems play an important role in fields such as chemistry, economy, population dynamics or even in transportation modeling where the state variables of the models are often physically constrained to be nonnegative \cite{Farina2000}. It is remarked that many non-positive systems can be transformed into the positive class either through invertible coordinates transformations or by using other approaches, where the distortion of the phase-space can be kept minimal in the region of interest \cite{Samardzija1989}.

Chemical Reaction Networks (CRNs) form a wide class of positive (or nonnegative) systems attracting significant attention not only among chemists but in numerous other  fields such as physics, or even pure and applied mathematics where nonlinear dynamical systems are considered \cite{Erdi1989}. Beside pure chemical reactions, CRNs are often used to model the dynamics of enzymatic systems \cite{Berberan-Santos2010}, intracellular processes, metabolic or cell signalling pathways \cite{Haag2005}.  The increasing interest towards reaction networks among mathematicians and engineers is clearly shown by recent tutorial and survey papers \cite{Son:2001, Angeli2009, Chellaboina2009}.

It is known from the so-called "fundamental dogma of chemical kinetics" that reaction networks with different graph structures and even with different sets of complexes might generate identical dynamical system models (i.e. sets of differential equations). This means that CRNs with structurally very different reaction mechanisms can show exactly the same behaviour in the state space that is usually the space of chemical specie concentrations.  However, many strong analysis results of chemical reaction network theory (CRNT) depend on the graph structure of the studied CRN. There is a clear need therefore to define and search for distinguished structures among the possible alternatives. The integration of logical expressions into mixed integer programming problems \cite{Raman1991, Bemporad1999} has opened the possibility to formulate the computation of certain reaction structures with advantageous properties as an optimization problem \cite{Szederkenyi2009b}.

Mixed Integer Nonlinear Programs (MINLPs) are the most general constrained optimization problems with a single objective. These problems can contain continuous and integer decision variables without any limitations to the form and complexity of the objective function or the constraints. As it is expected, the solution of these problems is rather challenging \cite{Floudas2005}. 
A special subset of optimization problems is the class of Mixed Integer Linear Programs (MILPs) where the objective function and the constraints are linear functions of the decision variables. Effective solvers have been developed for MILPs, although it is known that their solution is NP-hard. 
In the chemical and biochemical fields, efficient combinatorial optimization algorithms are widely applied e.g. in permanental polynomial computation \cite{Liang2008}, metabolic pathway construction, control analysis or metabolic network reconstruction \cite{Banga2008}. It is noted that the evolutionary approach can also be very successful in solving complex chemically originated optimization problems \cite{Kargar2009}.

In \cite{Szederkenyi2009b}, the notion of \textit{realization} was introduced for the unique definition of a reaction network, and a mixed integer linear programming (MILP)-based numerical procedure was proposed to compute sparse and dense realizations of mass-action reaction networks corresponding to the same mathematical model, solving important part of a problem that was raised almost 30 years ago in \cite{Hars1981}. The purpose of this paper is to present new results in the field of optimization based generation of reaction network structures.

\section{Basic notions and tools}
\subsection{Structural and dynamic description of CRNs obeying the mass action law}\label{subsec:MAK}
The overview in this subsection is largely based on \cite{Szederkenyi2009b}.
A CRN obeying the mass action law
 is a closed system under isothermal and isobaric conditions, where 
 chemical species $\mathbf{X}_i,~i=1,...,n$ take part in $r$ 
 chemical reactions. 
The concentrations of the species denoted by $x_i,~(i=1,...,n)$ form the state vector, i.e. $x_i=[\mathbf{X}_i]$. 
The \emph{elementary reaction steps} have the following form: 
\begin{equation} \label{Eq:irrevreakt}
 \sum_{i=1}^n \alpha_{ij} \mathbf{X}_{i}   \rightarrow
  \sum_{i=1}^n  \beta_{ij} \mathbf{X}_{i},~~j=1,...,r
\end{equation}
where $\alpha_{ij}$ is the so-called \textit{stoichiometric coefficient} 
 of component $\mathbf{X}_{i}$ in the $j$th reaction, and 
 $\beta_{i \ell}$ is the stoichiometric coefficient of the product 
 $\mathbf{X}_{\ell}$. The linear combinations of the species in eq. (\ref{Eq:irrevreakt}), namely $\sum_{i=1}^{n} \alpha_{ij}\mathbf{X}_{i}$ and $\sum_{i=1}^{n}
\beta_{ij}\mathbf{X}_{i}$ for $j=1,\dots,r$ are called the \textit{complexes} and are denoted by $C_1, C_2, \dots, C_m$.
Note that \textit{the stoichiometric coefficients are always nonnegative 
 integers in classical reaction kinetic systems}. 
The reaction rates of the 
 individual reactions can be described as 
\begin{equation} \label{Eq:MALrate} 
 \rho_j = k_j \prod_{i=1}^{n} [\mathbf{X}_{i}]^{\alpha_{ij}} = 
  k_j \prod_{i=1}^{n} x_i^{\alpha_{ij}}~~,~~j=1,...,r
\end{equation}
where $k_j>0$ is the \textit{reaction rate constant} 
 of the $j$th reaction. 

If the reactions $C_i\rightarrow C_j$ and $C_j\rightarrow C_i$ take place at the same time in a reaction network for some $i,j$ then this pair of reactions is called a reversible reaction (but it will be treated as two separate elementary reactions).

Similarly to \cite{Feinberg:79}, we can assign the following directed graph (see, e.g. \cite{Bang-Jensen2001}) to the reaction network \eqref{Eq:irrevreakt} in a straightforward way.
The directed graph $D=(V_d,E_d)$ of a reaction network consists of a finite nonempty set $V_d$ of vertices and a finite set $E_d$ of ordered pairs of distinct vertices called directed edges. The vertices correspond to the complexes, i.e. $V_d=\{C_1,C_2,\dots C_m\}$, while the directed edges represent the reactions, i.e. $(C_i,C_j)\in E_d$ if complex $C_i$ is transformed to $C_j$ in the reaction network. The reaction rate coefficients $k_j$ for $j=1,\dots,r$ in \eqref{Eq:MALrate} are assigned as positive weights to the corresponding directed edges in the graph. Where it is more convenient, the notation $k'_{ij}$ will be used for denoting the reaction rate coefficient corresponding to the reaction $C_i \rightarrow C_j$.

A set of complexes $\{C_1,C_2,\dots,C_k\}$ is a \textit{linkage class} of a reaction network if the complexes of the set are linked to each other in the reaction graph but not to any other complex \cite{Feinberg1987}. 
There are several possibilities to represent the dynamic equations of mass action systems (see, e.g. \cite{Feinberg:79}, \cite{Gorban2004},  or \cite{Craciun2008}). The most advantageous form for our purposes is the one that is used e.g. in Lecture 4 of \cite{Feinberg:79}, i.e.
\begin{equation}
\dot{x}=Y \cdot A_k\cdot \psi(x)\label{eq:Feinberg_desc}
\end{equation}
where $x\in\mathbb{R}^n$ is the concentration vector of the species, $Y\in\mathbb{R}^{n\times m}$ stores the stoichiometric composition of the complexes, $A_k\in\mathbb{R}^{m\times m}$ contains the information corresponding to the weighted directed graph of the reaction network, and $\psi:\mathbb{R}^n\mapsto\mathbb{R}^m$ is a monomial-type vector mapping defined by
\begin{equation}
\psi_j(x)=\prod_{i=1}^n x_i^{y_{ij}},~~~j=1,\dots,m
\end{equation} 
where $y_{ij}=[Y]_{ij}$. It is remarked that the numerical solution of kinetic differential equations can be a challenging task requiring advanced integration approaches \cite{Tsitouras2008}.
The exact structure of $Y$ and $A_k$ is the following. The $i$th column of $Y$ contains the composition of complex $C_i$, i.e. $Y_{ji}$ is the stoichiometric coefficient of $C_i$ corresponding to the specie $\mathbf{X}_j$. $A_k$ is a column conservation matrix (i.e. the sum of the elements in each column is zero) defined as
\begin{equation}
[A_k]_{ij}=\left\{
\begin{array}{ccc}
-\sum_{l=1}^m k'_{il}, & \text{if} & i=j \\
k'_{ji}, & \text{if} & i\ne j
\end{array}
\right.
\end{equation}
In other words, the diagonal elements $[A_k]_{ii}$ contain the negative sum of the weights of the edges starting from the node $C_i$, while the off-diagonal elements $[A_k]_{ij}$, $i\ne j$ contain the weights of the directed edges $(C_j,C_i)$ coming into $C_i$. Based on the above properties, it is appropriate to call $A_k$ the \textit{Kirchhoff matrix} of a reaction network.

To handle the exchange of materials between the environment and the reaction network, the so-called "zero-complex" can be introduced and used which is a special complex where all stoichiometric coefficients are zero i.e., it is represented by a zero vector in the $Y$ matrix (for the details, see, e.g. \cite{Feinberg:79} or \cite{Craciun2005}).

We can associate an $n$-dimensional vector with each reaction in the following way. For the reaction $C_i\rightarrow C_j$, the corresponding \textit{reaction vector} denoted by $h_k$ is given by
\begin{equation}
h_k=[Y]_{\cdot, j} - [Y]_{\cdot, i}\label{eq:reacto_vect}
\end{equation}
where $[Y]_{\cdot, i}$ denotes the $i$th column of $Y$. Similarly to reaction rate coefficients, whenever it is more practical, $h'_{ij}$ denotes the reaction vector corresponding to the reaction $C_i\rightarrow C_j$.

The \textit{rank} of a reaction network denoted by $s$ is defined as the rank of the vector set $H=\{h_1,h_2\dots,h_r \}$ where $r$ is the number of reactions. The elements of $H$ span the so-called \textit{stoichiometric subspace} denoted by $S$, i.e. $S=\text{span}\{h_1,h_2\dots,h_r \}$. The  positive \textit{stoichiometric compatibility class} containing a concentration $x_0$ is the following set \cite{Feinberg1987}:
\[
(x_0+S)\cap\mathbb{R}^n_+
\] 
where $\mathbb{R}^n_+$ denotes the positive orthant in $\mathbb{R}^n$.

The deficiency $d$ of a reaction network is defined as \cite{Feinberg:79, Feinberg1987}
\begin{equation}
d=m-l-s
\end{equation}
where $m$ is the number of complexes in the network, $l$ is the number of linkage classes and $s$ is the rank of the reaction network. 

A reaction network is called \textit{reversible}, if each of its reactions is a reversible reaction. A reaction network is called \textit{weakly reversible}, if each complex in the reaction graph lies on at least one directed cycle (i.e. if complex $C_j$ is reachable from complex $C_i$ on a directed path in the reaction graph, then $C_i$ is reachable from $C_j$ on a directed path). An important point of the well-known \textit{Deficiency Zero Theorem} \cite{Feinberg1987} says that the ODEs of a weakly reversible deficiency zero CRN are globally stable with a known logarithmic Lyapunov function for all positive values of the reaction rate coefficients. Therefore (among other realization problems) it is of interest whether we can find a (weakly) reversible deficiency zero kinetic realization of a nonnegative polynomial system. 

Using the notation $M=Y\cdot A_k$, eq. \eqref{eq:Feinberg_desc} can be written in the compact form
\begin{equation}
\dot{x}=M\cdot\psi(x)\label{eq:react_compact}
\end{equation}
The invariance of the nonnegative orthant for CRN dynamics is shown e.g. in \cite{Chellaboina2009}.

\subsection{Kinetic realizability of positive (nonnegative) polynomial systems}
An autonomous polynomial nonlinear system of the form 
\begin{align}
\dot{x}=f(x) \label{eq:nonlsys}
\end{align}
is called \textit{kinetically realizable} or simply \textit{kinetic}, if a mass action reaction mechanism given by eq. \eqref{eq:Feinberg_desc} can be associated to it that exactly realizes its dynamics, i.e. $f(x)=Y\cdot A_k \cdot \psi(x)$ where $\psi$ contains the monomials, matrix $Y$ has nonnegative integer elements and $A_k$ is a valid Kirchhoff matrix (see section \ref{subsec:MAK} for its properties). In such a case, the pair $(Y,A_k)$ will be called a \textit{realization} of the system \eqref{eq:react_compact} (note that $Y$ contains all information about the composition of the monomials in $\psi$ in the case of mass-action dynamics).  As it is expectable from linear algebra, the same polynomial system may have many parametrically and/or structurally different realizations. Thus, two CRNs will be called \textit{dynamically equivalent} if they realize the same polynomial system of the form \eqref{eq:nonlsys}. Therefore, CRN $\mathcal{A}$ will also be called a \textit{realization} of CRN $\mathcal{B}$, if $\mathcal{A}$ and $\mathcal{B}$ are dynamically equivalent.

The problem of kinetic realizability of polynomial vector fields was first examined and solved in \cite{Hars1981} where the constructive proof contains a realization algorithm that produces the directed graph of a possible associated mass action mechanism. It is important to remark here that the above mentioned realization algorithm typically produces high deficiency CRNs that are non-minimal in the sense that they usually contain more reactions and complexes than the minimal numbers that are necessary to realize the given kinetic polynomial system. According to \cite{Hars1981}, the necessary and sufficient condition for kinetic realizability is that all coordinates functions $f_i$ of the right hand side of \eqref{eq:nonlsys} must have the form
\begin{equation}
f_i(x) = -x_i g_i(x) + h_i(x),~i=1,\dots,n \label{react_real_form}
\end{equation}
where $g_i$ and $h_i$ are polynomials with nonnegative coefficients. 
\subsection{Mixed integer linear programming}
A special subset of optimization problems is the class of Mixed Integer Linear Programs (MILPs) where the objective function and the constraints are linear functions of the decision variables. 
A mixed integer linear program with $k$ variables (denoted by $w\in\mathbb{R}^k$) and $p$ constraints can be written as \cite{Nemhauser1988}:
\begin{align}
&\text{minimize}~c^Tw \nonumber\\
&\text{subject to:}\nonumber\\
&A_1w=b_1\nonumber\\
&A_2w\le b_2\label{eq:MILP_problem}\\
&l_i \le w_i \le u_i~\text{for}~i=1,\dots,k\nonumber\\
&w_j~\text{is integer for}~j\in I,~I\subseteq\{1,\dots,k\}\nonumber
\end{align}
where $c\in\mathbb{R}^k$, $A_1\in\mathbb{R}^{p_1\times k}$, $A_2\in\mathbb{R}^{p_2\times k}$, and $p_1+p_2=p$.

If all the variables can be real, then \eqref{eq:MILP_problem} is a simple linear programming problem that can be solved in polynomial time. However, if any of the variables is integer, then the problem becomes NP-hard. In spite of this, there exist a number of free (e.g. YALMIP or the GNU Linear Programming Kit) and commercial (such as CPLEX or TOMLAB) solvers that can efficiently handle many practical problems \cite{CPLEX2007, Loefberg2004, Makhorin2006}.

A propositional logic problem, where a statement denoted by $S$ must be proved to be true given a set of compound statements containing so-called literals $S_1,\dots,S_n$, can be solved by means of a linear integer program. For this, logical variables denoted by $\delta_i$ ($\delta_i\in\{0,1\}$) must be associated with the literals $S_i$. Then the original compound statements can be translated to linear inequalities involving the logical variables $\delta_i$ \cite{Raman1994, Bemporad1999}. 
\subsection{Computing CRN realizations with the minimal/maximal number of reactions as a MILP problem}\label{sec:dense_sparse}
For convenience, this subsection briefly summarizes the results of \cite{Szederkenyi2009b} without going into the details. The starting point is that a kinetic polynomial system of the form \eqref{eq:react_compact} is given with its parameters. This means that $M$ is known, the stoichiometrix matrix $Y$ is also known from the monomials of $\psi$, and we would like to determine the Kirchhoff matrix $A_k\in\mathbb{R}^{m\times m}$ that fulfils given requirements. 

The characteristics of the mass-action dynamics can be expressed in the form of the following equality and inequality constraints:
\begin{align}
& Y\cdot A_k = M \label{constr_1}\\
& \sum_{i=1}^m [A_k]_{ij}=0,~~~j=1,\dots,m \label{constr_2}\\
& [A_k]_{ij}\ge 0,~~i,j=1,\dots,m,~~i\ne j \\
& [A_k]_{ii}\le 0,~~i=1,\dots,m \label{constr_4}
\end{align}
where the decision variables are the elements of $A_k$. Clearly, constraints \eqref{constr_2}-\eqref{constr_4} express that we are searching for a valid Kirchhoff connection matrix. To make the forthcoming optimization problems computationally tractable, appropriate upper and lower bounds are introduced for the elements of $A_k$ as
\begin{align}
&0\le [A_k]_{ij} \le l_{ij},~~i,j=1,\dots,m,~~i\ne j \label{constr_5}\ \\
&l_{ii} \le [A_k]_{ii} \le 0,~~i=1,\dots,m \label{constr_6}.
\end{align}
In this problem set, we are searching for such $A_k$ that contains the minimal/maximal number of nonzero off-diagonal elements. For this, we introduce logical variables denoted by $\delta$ and construct the following compound statements
\begin{align}
\delta_{ij}=1 \leftrightarrow [A_k]_{ij}>\epsilon,~~i,j=1,\dots,m,~~i\ne j \label{compound01}
\end{align}
where the symbol "$\leftrightarrow$" represents "if and only if", and $0<\epsilon\ll 1$ (i.e. elements of $A_k$ below $\epsilon$ are treated as zero). Taking into consideration \eqref{constr_5}, statement \eqref{compound01} can be translated to the following linear inequalities (see, e.g. \cite{Bemporad1999})
\begin{align}
0\le [A_k]_{ij}-\epsilon \delta_{ij},~~i,j=1,\dots,m,~i\ne j \label{constr_ds_1}\\
0\le -[A_k]_{ij} + l_{ij}\delta_{ij},~~i,j=1,\dots,m,~i\ne j \label{constr_ds_2}
\end{align}
Now we are able to compute the realization containing the minimal/maximal number of reactions by minimizing/maximizing the objective function
\begin{align}
C_1(\delta)=\sum_{\begin{small}\begin{array}{c}i,j=1\\i\ne j\end{array}\end{small}}^m \delta_{ij}\label{ds_objfun}
\end{align}
The realizations of a reaction network containing the minimal and maximal number of reactions will be called the \textit{sparse} and \textit{dense} realizations, respectively \cite{Szederkenyi2009b}.
\subsection{A simple motivating example}
\begin{figure}
\centering
    \includegraphics[width=7cm]{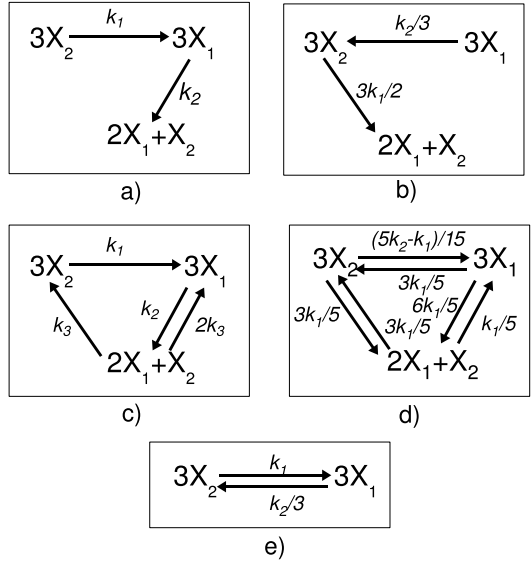}
    \caption{Dynamically equivalent reaction networks}
    \label{FIG:eq_reactions}
\end{figure}

Consider the simple reaction mechanism depicted in Fig. \ref{FIG:eq_reactions} a). It is easy to check that the reaction structures in Figs \ref{FIG:eq_reactions} b), c), d) and e) lead to the same dynamical description as the original structure a), namely
\begin{align}
\dot{x}_1 & = 3k_1 x_2^3 - k_2 x_1^3 \nonumber\\
\dot{x}_2 & = -3 k_1 x_2^3 + k_2 x_1^3, \label{eq:ex01}
\end{align}
with $k_1,k_2>0$, $5k_2>k_1$ (i.e. the CRNs in Fig. \ref{FIG:eq_reactions} are dynamically equivalent).
It is worth having a look at the structural properties of the different realizations of eq. \eqref{eq:ex01} shown in the subfigures. The realizations in Figs. \ref{FIG:eq_reactions}.a) and b) are irreversible, the structure in Fig. \ref{FIG:eq_reactions}.c) is weakly reversible, while the networks in Figs. \ref{FIG:eq_reactions}.d) and e) are fully reversible. The deficiencies of the first four realizations a)--d) are 1, while the deficiency of realization e) is zero. This means that both the weaker \textit{Deficiency one theorem} and the stronger \textit{Deficiency zero theorem} can be applied to all realizations a)--e), and this way to the dynamical system described by eq. \eqref{eq:ex01} (see \cite{Feinberg1987}). Shortly speaking, the Deficiency one theorem for such weakly reversible networks as c) says that its differential equations admit precisely one steady state in each positive stoichiometric compatibility class. Moreover, by applying the Deficiency zero theorem to realization e), we obtain the additional valuable fact that each steady state of \eqref{eq:ex01} is asymptotically stable within the corresponding positive stoichiometric compatibility class with the Lyapunov function:
\begin{align}
V(x)=\sum_{i=1}^2 x_i \left(\ln\left(\frac{x_i}{x_i^*} \right) -1\right) + x_i^*,
\end{align}
where $x^*$ denotes the equilibrium point of \eqref{eq:ex01} corresponding to the given stoichiometric compatibility class.

First of all, the above example shows very transparently that important structural properties such as deficiency, reversibility or weak reversibility are not encoded uniquely in the polynomial differential equations of a kinetic system. Secondly, it is definitely of interest to develop computational tools to search for realizations with such properties that are useful in the dynamical analysis of given kinetic polynomial systems or CRNs.

\section{Properties of dense realizations and its consequences}
The main result of this section is that the dense realization of a CRN is structurally unique if the set of possible complexes is fixed. We recall that the realizations of a reaction network containing the minimal and maximal number of nonzero reaction rate coefficients are called the \textit{sparse} and \textit{dense} realizations.
\subsection{The uniqueness of the structure of dense realizations}
Firstly, we state the following result.
\begin{theorem}\label{th:01}
If a set of kinetic differential equations denoted by $\Sigma$ is given with matrices $M$ and $Y$, then the directed unweighted graph of any realization of $\Sigma$ must be a subgraph of the directed unweighted graph of the dense realization.
\end{theorem}
\begin{proof}
The proof is based on the following elementary fact of linear algebra. Consider an inhomogenous set of linear equations:
\begin{align}
Ax & = b \label{eq:inhom}
\end{align}
If $p$ is any specific solution of \eqref{eq:inhom} then the entire solution set of \eqref{eq:inhom} can be given as
\begin{align}
\{p+v~|~v~\text{is any solution of}~Ax=0 \}\label{eq:lin_solution}
\end{align}
The matrix equation $Y\cdot A_k = M$ (see eqs. \eqref{eq:Feinberg_desc} and \eqref{eq:react_compact}) obviously defines $m$ sets of linear equations of the form
\begin{align}
Y\cdot [A_k]_{\cdot,i}=[M]_{\cdot,i},~~i=1,\dots,m \label{eq:react_lineq}
\end{align}
where the unknown is $[A_k]_{\cdot,i}$ (i.e., the $i$th column of $A_k$). For any $i$, let us assume that $p=[A_k]_{\cdot,i}$ is a dense solution of \eqref{eq:react_lineq} i.e., it contains the maximal possible number of nonzero elements. Let $l$ denote the number of nonzero elements in $p$. If $l=m$ then the theorem is trivial, so from now on we assume that $l<m$. Let us assume furthermore that $p'\ne p$ is also a solution of \eqref{eq:react_lineq} and let $(j_1,\dots,j_q)$ denote the indices where $p(j_k)= 0$ while $p'(j_k)\ne 0$ for $k=1,\dots q$. With $q>0$, this means that the directed unweighted reaction graph defined by $p'$ is not a subgraph of the directed unweighted reaction graph defined by $p$. Then, according to \eqref{eq:lin_solution}, $p'$ can be written as
\begin{align}
p'=p+v,
\end{align}
where $Y\cdot v = 0$. According to our assumption, $v\ne 0$, and necessarily, $v(j_1)\ne 0, \dots, v(j_q)\ne 0$.    
Let $l'$ denote the number of nonzeros in $p'$. Since $l' < l$ (because $p$ is a dense solution), there must exist indices $(h_1,\dots,h_z)$ disjoint from $(j_1,\dots,j_q)$ with $z \ge q$ such that $v(h_k)=-p(h_k)\ne 0$ for $k=1,\dots,z$. Then for any $\lambda\in\mathbb{R}$, $p''=p+\lambda\cdot v$ is also a solution of \eqref{eq:react_lineq}, and $\lambda$ can always be chosen such that $p''$ contains more nonzero elements than $p$, which is clearly a contradiction. 
\end{proof}
We note that we did not use the further restriction that $[A_k]_{\cdot,i}$ is an appropriate column of a Kirchhoff matrix, but this was not needed for the proof. Now we easily obtain our following result about the uniqueness of the dense realization.
\begin{theorem}\label{th:02}
If a set of kinetic differential equations denoted by $\Sigma$ is given with matrices $M$ and $Y$, then the directed graph structure of its dense realization is unique.
\end{theorem}
\begin{proof}
The proof is the special case of the proof of Theorem \ref{th:01} with $l'=l$ and $q=z$.
\end{proof}
\subsection{Important consequences and special cases}
The following remarks contain some important immediate consequences and additions to Theorems \ref{th:01} and \ref{th:02}.
\begin{itemize}
\item[\textbf{R1}] According to Theorems \ref{th:01} and \ref{th:02}, dense realizations give a unique "superstructure" for a CRN in the sense that the reactions of any realization of a CRN must form a subset of the reactions of the dense realization if the set of possible complexes is given. In other words, reactions that are not present in the dense realization cannot appear in any other realization. 

\item[\textbf{R2}] Obviously, dense realizations are parametrically not unique. There may exist several dense realizations for a CRN with different reaction rate constants (weights) but always with the same graph structure.

\item[\textbf{R3}] The graph structure of a CRN with a given set of complexes is unique if and only if the graph structures of its sparse and dense realizations are identical.

This fact is easy to see: If the structures of the dense and sparse realizations are identical, then it directly follows that the graph structure of the CRNs is unique, since the only possible unique structure is determined by the dense realization (that is the sparse realization at the same time). In other words, any realization of the CRN can contain neither more nor less reactions than the dense realization does, the structure of which is unique. If the graph structure of the CRN is unique, then it trivially implies that the structures of the dense and sparse realizations are identical.

\item[\textbf{R4}] The dense realization of a CRN is not only a theoretical construction but it can be
practically determined using well-formulated numerical procedures that are treatable even in the case of several hundred complexes and species (see, e.g. \cite{Floudas1995, Raman1994}).

\item[\textbf{R5}] Sparse realizations of CRNs are structurally not unique, there may exist several sparse realizations for a given CRN with different graph structures (see later in subsection \ref{sec:sparse_nonunique}).
\end{itemize}

\section{Transforming additional constraints corresponding to preferred CRN properties into linear inequalities}
This section presents some further answers to the open problems originally set in \cite{Hars1981} from an optimization point of view.
\subsection{Computing realizations with the minimal/maximal number of complexes}\label{sec:compminmax}
In this section, the detailed MILP formalism will be presented for computing CRN realizations that contain the minimal/maximal number of complexes from a predefined complex set. 

Let us assume again that the set of feasible complexes is a'priori given with matrix $Y$. The constraints written in eqs. \eqref{constr_1}-\eqref{constr_6} corresponding to the characteristics of mass-action dynamics are used here again without change. Then, the minimization or maximization of the number of non-isolated complexes in the reaction graph is based on the following simple observation. A complex disappears from the reaction network's graph, if both the corresponding column and row in $A_k$ contain only zeros. This means that no directed edges start from or point to this complex in the graph and therefore it becomes an isolated
vertex that can be omitted.

For the optimization, $m$ boolean variables denoted by $\delta_i$, $i=1,\dots,m$ are introduced. Using these boolean variables, the following compound statements are introduced:
\begin{align}
\delta_i=1 \leftrightarrow \sum_{\begin{array}{c}\scriptstyle j_1=1\\\scriptstyle j_1\ne i \end{array}}^m [A_k]_{i,j_1} +
\sum_{\begin{array}{c}\scriptstyle j_2=1\\ \scriptstyle j_2\ne i \end{array}}^m [A_k]_{j_2,i}>0,~~i=1 \dots, m \label{comp_compminmax1}
\end{align}
Eq. \eqref{comp_compminmax1} means that the value of $\delta_i$ is 1 if and only if there is at least incoming/outgoing directed edge in the reaction graph to/from the $i$th complex.
For practical computations, the statement \eqref{comp_compminmax1} is modified as follows:
\begin{align}
\delta_i=1 \leftrightarrow \sum_{\begin{array}{c}\scriptstyle j_1=1\\ \scriptstyle j_1\ne i \end{array}}^m [A_k]_{i,j_1} +
\sum_{\begin{array}{c}\scriptstyle j_2=1\\ \scriptstyle j_2\ne i \end{array}}^m [A_k]_{j_2,i}>\epsilon,~~i=1 \dots, m\label{comp_compminmax2}
\end{align}
where again $0<\epsilon\ll 1$ (see, eq. \eqref{compound01}). Using the bound constraints \eqref{constr_5}-\eqref{constr_6}, the linear inequalities corresponding to \eqref{comp_compminmax2} are the following
\begin{small}
\begin{align}
&0\le \sum_{\begin{array}{c}\scriptstyle j_1=1\\ \scriptstyle j_1\ne i \end{array}}^m [A_k]_{i,j_1} + 
\sum_{\begin{array}{c}\scriptstyle j_2=1\\ \scriptstyle j_2\ne i \end{array}}^m [A_k]_{j_2,i}-\epsilon \delta_i,~~i=1,\dots,m \\
&0 \le - \sum_{\begin{array}{c}\scriptstyle j_1=1\\ \scriptstyle j_1\ne i \end{array}}^m [A_k]_{i,j_1} -
\sum_{\begin{array}{c}\scriptstyle j_2=1\\ \scriptstyle j_2\ne i \end{array}}^m [A_k]_{j_2,i} + \epsilon +
\left(\sum_{\begin{array}{c}\scriptstyle j_1=1\\ \scriptstyle j_1\ne i \end{array}}^m l_{i j_1} + \sum_{\begin{array}{c}\scriptstyle j_2=1\\ \scriptstyle j_2\ne i \end{array}}^m l_{j_2 i} - \epsilon\right)\cdot\delta_i,~~ i=1,\dots, m
\end{align}
\end{small}
Now, the objective function to be minimized or maximized can be written as
\begin{align}
C_2(\delta)=\sum_{i=1}^m \delta_i
\end{align}
In contrast to the algorithm summarized in section \ref{sec:dense_sparse}, minimizing/maximizing the number of non-isolated complexes is not straightforward to parallelize (see also \cite{Szederkenyi2009b}). However, the number of integer variables in this case is only $m$, compared to $m^2-m$ when minimizing/maximizing the number of reactions.
\subsection{Computing reversible realizations}\label{sec:reversible}
Here, the basic constraints \eqref{constr_1}-\eqref{constr_6} expressing the properties of mass action dynamics and lower and upper bounds for the reaction rate coefficients will be used again for the optimization. To distinguish between zero and nonzero reaction rate coefficients, a small positive scalar $\epsilon$ is applied again, similarly to the previous case described in section \ref{sec:compminmax}.

The additional constraint for the full reversibility of the CRN structure is not difficult to formulate as 
\begin{align}
[A_k]_{i,j}>\epsilon_2 \leftrightarrow [A_k]_{j,i}>\epsilon_2,~~\forall i>j. \label{constr_reversibility}
\end{align}
where $\epsilon_2$ is a positive threshold value such that $\epsilon<\epsilon_2$. The linear inequalities equivalent to \eqref{constr_reversibility} can be written as
\begin{align}
0 \le (\epsilon_2-\epsilon) - [A_k]_{ij} + (l_{ij}-\epsilon_2)\cdot\delta^{(1)}_{ij},~~\forall i>j \label{revconstr_1}\\
0 \le (\epsilon_2-\epsilon) - [A_k]_{ji} + (l_{ji}-\epsilon_2)\cdot\delta^{(1)}_{ij},~~\forall i>j \label{revconstr_2}\\
0 \le [A_k]_{ij} - \epsilon_2\cdot\delta^{(1)}_{ij},~~\forall i>j \label{revconstr_3}\\
0 \le [A_k]_{ji} - \epsilon_2\cdot\delta^{(1)}_{ij},~~\forall i>j \label{revconstr_4}
\end{align}
where $l_{ij}$ is the upper bound for $[A_k]_{ij}$ as it is introduced in eq. \eqref{constr_5}. Furthermore, $\frac{m(m-1)}{2}$ integer variables are introduced for the representation of the reversibility constraint that are denoted by $\delta^{(1)}_{ij}$, $\forall i>j$. 

In order to exclude reaction rate coefficients between $\epsilon$ and $\epsilon_2$, and to obtain a numerically stable solution, the following additional constraints in the form of a compound statement are introduced
\begin{align}
[A_k]_{ij}<\epsilon~~\text{OR}~~[A_k]_{ij}>\epsilon_2+\gamma, \label{excl_coeffs}
\end{align}
where $\gamma$ is a small positive threshold value that is in the same order of magnitude as $\epsilon_2$. The set of inequalities equivalent to \eqref{excl_coeffs} is given by
\begin{align}
0 & \le \delta^{(2)}_{ij},~~i\ne j \label{revconstr_5}\\
0 & \le l_{ij} - [A_k]_{ij}-(l_{ij}-\epsilon)\cdot\delta^{(3)}_{ij},~~i\ne j\\
0 & \le [A_k]_{ij}-(\epsilon_2+\gamma)\cdot\delta^{(4)}_{ij},~~i\ne j \\
0 & \le -\delta^{(2)}_{ij} + \delta^{(3)}_{ij} + \delta^{(4)}_{ij},~~i\ne j\\
0 & \le \delta^{(2)}_{ij} - \delta^{(3)}_{ij},~~i\ne j \\
0 & \le \delta^{(2)}_{ij} - \delta^{(4)}_{ij},~~i\ne j \label{revconstr_10} 
\end{align}
where $\delta^{(2)}$, $\delta^{(3)}$ and $\delta^{(4)}$ represent altogether $3(m^2-m)$ integer variables.

It is remarked that the inequalities \eqref{revconstr_1} -- \eqref{revconstr_4} and \eqref{revconstr_5} -- \eqref{revconstr_10} express only constraints and no objective function is associated to reversibility in itself. However, the reversibility constraints can be easily combined with the minimization/maximization of either the number of reactions or that of the non-isolated complexes, still in the framework of mixed integer linear programming. Moreover, the strict reversibility constraint can be  modified into the minimization/maximization of reversible reactions in a straightforward way. It is emphasized finally, that the constraints presented in this subsection together with an appropriate MILP solver are suitable for deciding whether a reversible realization exists for a given CRN or not.
\section{Examples}
For the examples described in this section, the YALMIP modeling tool was applied under the MATLAB computational environment \cite{Loefberg2004} using both the freely available GLPK \cite{Makhorin2006} and the commercial CPLEX \cite{CPLEX2007} solvers. 
\subsection{Non-uniqueness of sparse realizations}\label{sec:sparse_nonunique}
\begin{figure}[h!]
\centering
    \includegraphics[width=10cm]{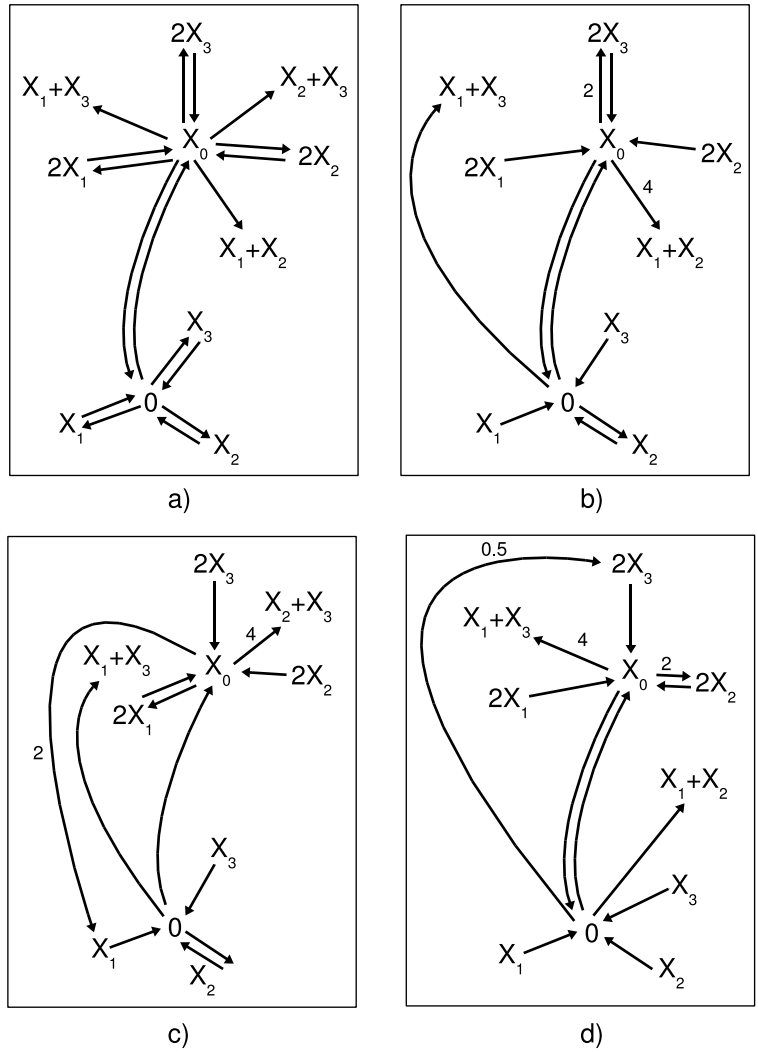}
    \caption{Original reaction network and its three different sparse realizations. Only reaction rates different from 1 are indicated.}
    \label{FIG:react_3_01}
\end{figure}

For the illustration of several structurally different sparse realizations of a reaction network, let us recall a literature example that was originally published in \cite{Craciun2008}. The original CRN with all reaction rate coefficients equal to 1 is shown in Fig. \ref{FIG:react_3_01} a). The CRNs in Figs. \ref{FIG:react_3_01} b) and c) were obtained by using the parallel and non-parallel version of the method described in \cite{Szederkenyi2009b} and summarized in section \ref{sec:dense_sparse}, respectively, using the GLPK MILP solver. The network shown in Fig. \ref{FIG:react_3_01} d) was computed using the CPLEX solver, using a non-parallel approach. These results show that additional constraints in the optimization procedure may be used to select the required sparse realization from the set of possible alternatives.

\subsection{Motivating example continued}\label{sec:mot_ex_cont}
Consider again the reaction network shown in Fig. \ref{FIG:eq_reactions} a) with parameters $k_1=1$, $k_2=2$. The matrices characterizing the CRN realization are the following.
\begin{align}
& Y=\begin{gmatrix}{rrr}
0 & 3 & 2 \\
3 & 0 & 1
\end{gmatrix},~~A_k=
\begin{gmatrix}{rrr}
-1 & 0 & 0 \\
1 & -2 & 0 \\
0 & 2 & 0
\end{gmatrix}
\end{align}
\begin{align}
& M=Y\cdot A_k=
\begin{gmatrix}{rrr}
3 & -2 & 0 \\
-3 & 2 & 0
\end{gmatrix}\label{ex:motivating}
\end{align}
\subsubsection{Computing a realization with the minimal number of complexes}
Finding a realization with the minimal number of complexes using the method described in section \ref{sec:compminmax} with parameters $l_{ij}=100$ $\forall i,j$ and $\epsilon=10^{-8}$ gives the following result:
\begin{align}
A_k^{(2)}=\begin{gmatrix}{rrr}
-1 & 0.6667 & 0 \\
1 & -0.6667 & 0 \\
0 & 0 & 0
\end{gmatrix}
\end{align}
It's straightforward to check that $M=Y\cdot A_k^{(2)}$. Here, $A_k^{(2)}$ gives a deficiency 0 structure that is shown in Fig. \ref{FIG:eq_reactions} e). 
\subsubsection{Computing a dense reversible realization}
If we search for a reversible realization given by eq. \ref{ex:motivating} that contains the maximal number of nonzero reaction rate coefficients (i.e. a dense reversible realization), we have to combine constraints \eqref{constr_1}-\eqref{constr_6}, \eqref{revconstr_1}-\eqref{revconstr_4}, \eqref{revconstr_5}-\eqref{revconstr_10} and  \eqref{constr_ds_1}-\eqref{constr_ds_2}, and maximize the objective function \eqref{ds_objfun}. Using the parameters $\epsilon=10^{-8}$, $\epsilon_2=0.05$, $\gamma=0.01$ we obtain a fully reversible structure given by the following Kirchhoff matrix
\begin{align}
A_k^{(3)}=\begin{gmatrix}{rrr}
 -1.0200  &  0.6467 &  33.3333\\
    0.9600 &  -0.7067 &  66.6667\\
    0.0600  &  0.0600 & -100.0000
\end{gmatrix}
\end{align}
which gives a deficiency 1 structure shown in Fig. \ref{FIG:eq_reactions}.d. Again, it's clear that $Y\cdot A_k=Y\cdot A_k^{(3)}$.
\subsection{Equivalent reversible realization of an irreversible reaction network}
\begin{figure}[h!]
\centering
    \framebox{\includegraphics[width=11cm]{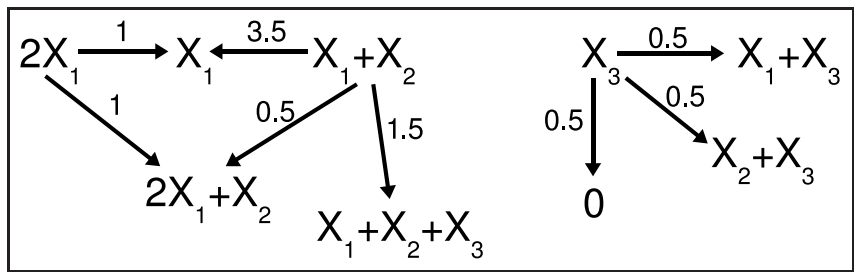}}
    \caption{Irreversible reaction network with a deficiency of 4}
    \label{FIG:react_4_01}
\end{figure}

Let us start from the reaction network that is depicted in Fig. \ref{FIG:react_4_01}. This network contains 9 complexes, 2 linkage classes and 8 irreversible reaction steps. The rank of the stoichiometric subspace is 3, therefore the deficiency of the network is 4. The matrices characterizing the network are given by
\begin{align}
Y=\begin{gmatrix}{lllllllll}
     2   &  1  &   1  &   2  &   0  &   1   &  0  &   1   &  0\\
     0  &   0  &   1  &   1  &   0  &   0  &   1  &   1  &   0\\
     0  &   0  &   0   &  0  &   1  &   1  &   1  &   1  &   0
\end{gmatrix},\\
A_k=\begin{gmatrix}{rrrrrrrrr}
   -2     &    0   &  0   &      0   &      0   &      0   &      0   &      0    &     0\\
    1     &    0  &  3.5    &     0  &       0   &      0   &      0   &      0    &     0\\
    0    &   0 &  -5.5    &     0     &    0     &    0     &    0     &    0     &    0\\
    1    &   0  &  0.5   &      0     &    0     &    0     &    0     &    0     &    0\\
         0   &      0    &     0    &     0 &  -1.5    &     0   &      0   &      0   &      0\\
         0     &    0     &    0    &     0  &  0.5    &     0   &      0     &    0     &    0\\
         0     &    0     &    0    &     0  &  0.5    &     0    &     0     &    0     &    0\\
         0     &    0    1.5     &    0      &   0     &    0    &     0     &    0     &    0\\
         0     &    0      &   0    &     0   & 0.5     &    0     &    0     &    0     &    0
\end{gmatrix}.
\end{align}
Running the algorithm described in section \ref{sec:reversible} with parameters $\epsilon=10^{-8}$, $\epsilon_2=0.05$, $\gamma=0.01$, where the objective function to be minimized was the number of nonzero reaction rate coefficients, gave the following Kirchhoff matrix:
\begin{align}
A_k'=
\begin{gmatrix}{rrrrrrrrr}
   -1     &    0   &  2   &      0   &      0   &      0   &      0   &      0    &     0\\
    0     &    0  &  0    &     0  &       0   &      0   &      0   &      0    &     0\\
    1    &   0 &  -3.5    &     0     &    0.5     &    0     &    0     &    0     &    0\\
    0   &   0  &  0   &      0     &    0     &    0     &    0     &    0     &    0\\
         0   &      0    &     1.5    &     0 &  -0.5    &     0   &      0   &      0   &      0\\
         0     &    0     &    0    &     0  &  0    &     0   &      0     &    0     &    0\\
         0     &    0     &    0    &     0  &  0    &     0    &     0     &    0     &    0\\
         0     &    0  &  0     &    0      &   0     &    0    &     0     &    0     &    0\\
         0     &    0      &   0    &     0   &  0     &    0     &    0     &    0     &    0
\end{gmatrix}
\end{align}
It is again easy to verify that
\begin{align}
Y\cdot A_k=Y\cdot A_k' = 
\begin{gmatrix}{rrrrrrrrr}
 -1    &     0  &  0.5   &      0 &   0.5   &      0   &      0  &       0   &      0\\
    1    &     0  & -3.5   &      0  &  0.5    &     0    &     0    &     0     &    0\\
         0   &      0  &  1.5   &      0  & -0.5   &      0   &      0    &     0   &      0
\end{gmatrix}
\end{align}
The above result implies that the deficiency zero theorem can be applied to the dynamics of the original irreversible reaction network shown in Fig. \ref{FIG:react_4_01}. Moreover, due to the existence of a deficiency 0 reversible realization with linearly independent reaction-pairs, the dynamics of the reaction networks exhibit a dissipative Hamiltonian structure as it was shown in \cite{Otero-Muras2008a}.
\begin{figure}[h!]
\centering
    \framebox{\includegraphics[width=6cm]{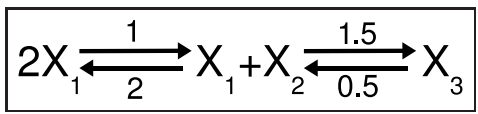}}
    \caption{Zero deficiency reversible reaction network dynamically equivalent to the one shown in Fig. \ref{FIG:react_4_01}}
    \label{FIG:react_4_02}
\end{figure}
\section{Conclusions}
Different possible realizations of dynamically equivalent CRNs have been studied in this paper with the help of mixed integer linear programming. The main contributions of the paper can be summarized as follows.
Firstly, it has been shown that the structure of a so-called dense realization of a given CRN is unique, and the structure of any other realization is the subgraph of the dense realization if the set of complexes is given. By computing a possible sparse realization, it is also possible to test numerically, whether the structure of a CRN is unique or not. Secondly, a method has been given for finding a CRN realization with the minimal number of complexes (from within a predefined set) in the framework of MILP. Finally, the numerically feasible constraints (linear (in)equalities) for determining reversible realizations of CRNs have been presented.  
The theoretical findings have been illustrated on examples. The results clearly show the power of linear programming combined with propositional logic for determining preferred realizations of reaction kinetic systems.

\section*{Ackowledgements}
This research work has been partially supported by the Hungarian Scientific Research Fund
 through grant no. K67625 and by the Control Engineering Research Group of the Budapest
 University of Technology and Economics. Gábor Szederkényi is a grantee of the Bolyai János Research 
Scholarship of the Hungarian Academy of Sciences.

\bibliography{react_MATCH}

\end{document}